\documentclass{ws-ijmpe}
\begin{document}


\markboth{M. G\'o\'zd\'z, F. \v Simkovic, and W. A.
  Kami\'nski}{Majorana neutrino magnetic moment}

%
\catchline{}{}{}{}{}
%

\title{MAJORANA NEUTRINO MAGNETIC MOMENTS}

\author{Marek G\'o\'zd\'z and Wies{\l}aw A. Kami\'nski}

\address{
Department of Theoretical Physics, Maria Curie-Sk{\l}odowska University,
Lublin, Poland \\
mgozdz@kft.umcs.lublin.pl \\ 
kaminski@neuron.umcs.lublin.pl}

\author{Fedor \v Simkovic}

\address{Department of Nuclear Physics, Comenius University, Bratislava,
Slovakia \\
simkovic@fmph.uniba.sk}

\maketitle

\begin{history}                %
\received{(xx October 2005)}	%
\end{history}               	%

\begin{abstract}
  The presence of trilinear $R$-parity violating interactions in the
  MSSM lagrangian leads to existence of quark--squark and
  lepton--slepton loops which generate mass of the neutrino. By
  introducing interaction with an external photon the magnetic moment is
  obtained. We derive bounds on that quantity being around one order of
  magnitude stronger than those present in the literature.
\end{abstract}


Thanks to more than 40 years of intensive work oscillations of
neutrinos are treated as a~well established experimental fact. Up to our
best knowledge this situation implies that neutrinos are massive
particles and therefore the Standard Model (SM) of elementary particles
and interactions should be extended. All such attempts to go beyond SM
are called non-standard physics and include supersymmetry (SUSY),
theories of grand unification (GUT), extra dimensions and others.

The problem of generation of very small neutrino masses has recently
received a~great deal of attention. There is, among others, a~mechanism
that relies on the $R$-parity violation in supersymmetric Standard
Model. The $R$-parity is defined as $R=+1$ for ordinary particles and
$R=-1$ for their supersymmetric partners. $R$-parity conservation makes
life easier because SUSY particles cannot decay into ordinary particles
and vice-versa. Theoretically, however, nothing motivates such behavior,
therefore many models consider violation of $R$-parity. This implies, in
turn, non-conservation of lepton and baryon numbers, which opens the
possibility for exotic nuclear processes to appear.

We will use the minimal supersymmetric standard model (MSSM), with
supersymmetry broken by supergravity (SUGRA).\cite{mg-mssm} The model
is fully described by the following superpotential and lagrangian. The
$R$-parity conserving part of the superpotential has the form
\begin{eqnarray}
  W^{MSSM} &=& \epsilon_{ab} [(\mathbf{Y}_E)_{ij} L_i^a H_1^b \bar E_j
    + (\mathbf{Y}_D)_{ij} Q_i^{ax} H_1^b \bar D_{jx} \nonumber \\
    &+& (\mathbf{Y}_U)_{ij} Q_i^{ax} H_2^b \bar U_{jx} + \mu H_1^a H_2^b ],
\end{eqnarray}
while its $R$-parity violating part reads
\begin{eqnarray}
  W^{RpV} &=& \epsilon_{ab}\left[
    \frac{1}{2} \lambda_{ijk} L_i^a L_j^b \bar E_k
    + \lambda'_{ijk} L_i^a Q_j^{xb} \bar D_{kx} \right] \nonumber \\
  &+& \frac{1}{2}\epsilon_{xyz} \lambda''_{ijk}\bar U_i^x\bar
  D_j^y \bar D_k^z + \epsilon_{ab}\kappa^i L_i^a H_2^b.
\end{eqnarray}
Here {\bf Y}'s are 3$\times$3 Yukawa matrices, $L$ and $Q$ stand for
lepton and quark left-handed $SU(2)$ doublet superfields while $\bar E$,
$\bar U$ and $\bar D$ denote the right-handed lepton, up-quark and
down-quark $SU(2)$ singlet superfields, respectively. $H_1$ and $H_2$
mean two Higgs doublet superfields. We have introduced color indices
$x,y,z = 1,2,3$, generation indices $i,j,k=1,2,3$ and the SU(2) spinor
indices $a,b,c = 1,2$.

The introduction of $R$-parity violation implies the existence of lepton
or baryon number violating processes, like the unobserved proton decay
and neutrinoless double beta decay ($0\nu2\beta$). Fortunatelly one may
keep only one type of terms and it is not necessary to have both
non-zero at one time. In order to get rid of too rapid proton decay and
to allow for lepton number violating processes it is customary to set
$\lambda''=0$.

We present here for completeness the mass term for scalar particles:
\begin{eqnarray}
{\cal L}^{mass} &=& \mathbf{m}^2_{H_1} h_1^\dagger h_1 +
                    \mathbf{m}^2_{H_2} h_2^\dagger h_2 +
     q^\dagger \mathbf {m}^2_Q q + l^\dagger \mathbf {m}^2_L l \nonumber \\
&+&  u \mathbf {m}^2_U u^\dagger + d \mathbf {m}^2_D d^\dagger +
     e \mathbf {m}^2_E e^\dagger,
\end{eqnarray}
soft gauginos mass term ($\alpha=1,...,8$ for gluinos):
\begin{equation}
  {\cal L}^{gaug.} = \frac12 \left( 
  M_1 \tilde{B}^\dagger \tilde{B} + 
  M_2 \tilde{W_i}^\dagger \tilde{W^i} +
  M_3 \tilde{g_\alpha}^\dagger \tilde{g^\alpha} + h.c.\right ),
\end{equation}
as well as the SUGRA mechanism, by introducing the soft supersymmetry
breaking Lagrangian
\begin{eqnarray}
  {\cal L}^{soft} &=& \epsilon_{ab} [(\mathbf{A}_E)_{ij} l_i^a h_1^b \bar e_j
    + (\mathbf{A}_D)_{ij} q_i^{ax} h_1^b \bar d_{jx} \nonumber \\
    &+& (\mathbf{A}_U)_{ij} q_i^{ax} h_2^b \bar u_{jx} + B \mu h_1^a h_2^b +
    B_2 \epsilon_i l_i^a h_2^b],
\end{eqnarray}
where lowercase letters stand for scalar components of respective chiral
superfields, and 3$\times$3 matrices {\bf A} as well as $B\mu$ and $B_2$
are the soft breaking coupling constants.

%
\begin{figure*}[b]
  \centerline{
    \includegraphics[width=0.45\textwidth]{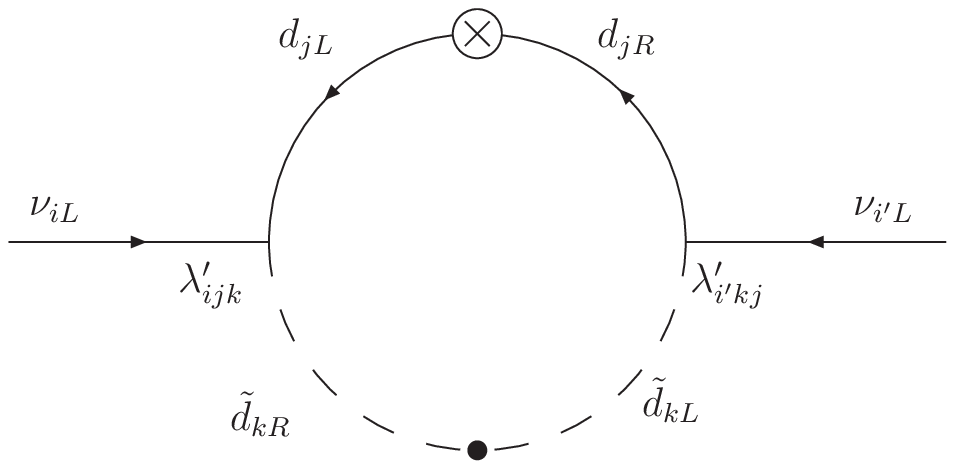}
    \includegraphics[width=0.45\textwidth]{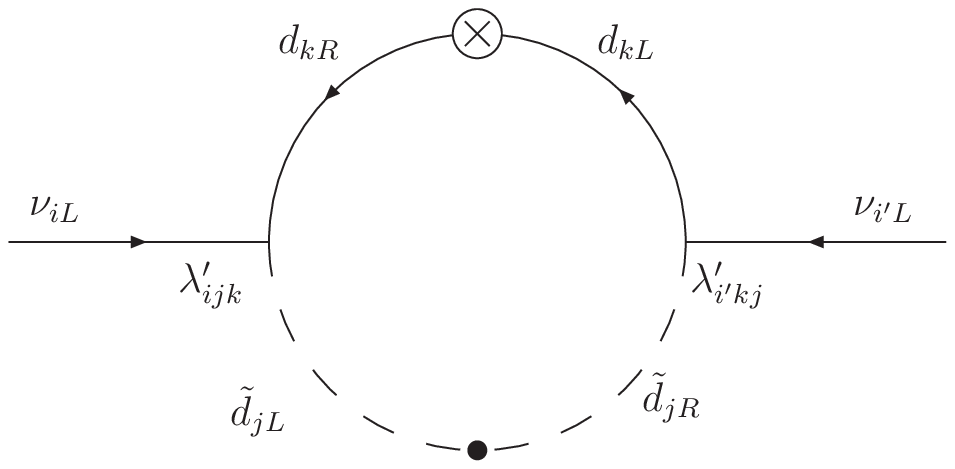}}
  \caption{\label{mg-fig1} Feynman diagrams leading to Majorana neutrino
    masses. By attaching photon to the internal line of the loop one
    gets the neutrino magnetic moment.}
\end{figure*}
%

The loops showed in Fig.~\ref{mg-fig1} lead to Majorana neutrino mass
term.\cite{mg-loops,mg-proc} Detailed calculation gives for the
quark--squark loop the following contribution:
\begin{equation} 
  M_{ii'}^q = {3 \over 16 \pi^2} \sum_{jk} \lambda'_{ijk}
  \lambda'_{i'kj} \left[ m_{q^j} {\sin(2\theta^k) \over 2}
    f(x_2^{jk},x_1^{jk}) + (j \leftrightarrow k) \right ],
\label{eq:mg-mqq}
\end{equation}
where $f(a,b) = \log(a)/(1-a) - \log(b)/(1-b)$, $m_{d^j}$ is $j$-th
generation down quark mass, $\theta^k$ is the squark mixing angle
between the $k$-th squark mass eigenstates $M_{\tilde q^k_{1,2}}$, and
$x_{1,2}^{jk} = m_{d^j}^2 / M_{\tilde q^k_{1,2}}^2$.

A~similar contribution comes from loops containing lepton--slepton
pairs. It reads:
\begin{equation}
  M_{ii'}^\ell = {1 \over 16 \pi^2} \sum_{jk} \lambda_{ijk}
  \lambda_{i'kj} \left[ m_{e^j} {\sin(2\phi^k) \over 2}
    f(y_2^{jk},y_1^{jk}) + (j \leftrightarrow k) \right ],
\label{eq:mg-mll}
\end{equation}
where all the quantities are defined in complete analogy with the
previous case, by replacing squarks with sfermions and quarks with
fermions. The factor 3 in Eq.~(\ref{eq:mg-mqq}) comes from summation
over three colors of quarks and therefore it is absent in the case of
fermions.

The left-hand sides of Eqs.~(\ref{eq:mg-mqq}) and (\ref{eq:mg-mll}) may
be obtained by considering neutrino oscillations phenomenology or by
using data from neutrinoless double beta decay experiments. The
half-life of this exotic process depends on the effective neutrino mass
$\langle m_\nu \rangle = \sum_{i=1}^3 |U_{ei}|^2 m_i$, where $U$ is the
neutrino mixing matrix and $m_i$ are neutrino mass eigenstates. There is
no firm experimental evidence for observation of the $0\nu2\beta$
process. The conservative constraint coming from the Heidelberg--Moscow
experiment is for now $T_{1/2}^{0\nu} \ge 1.9 \times 10^{25}$ years. By
assuming nuclear matrix elements of Ref.~\cite{mg-fedor} this value
translates into $\langle m_\nu \rangle \le 0.55$ eV, which in turn
implies the neutrino mass matrix in the form
\begin{eqnarray}
|M^{HM}| \le 
  \left (
  \begin{array}{ccc}
    0.55 & 0.78 & 0.66 \\
    0.78 & 0.56 & 0.69 \\
    0.66 & 0.69 & 0.81 
  \end{array}
  \right ) \rm{ eV}.
\end{eqnarray}
The entries of $M^{HM}$ may be used as the left-hand side of
Eqs.~(\ref{eq:mg-mqq}) and (\ref{eq:mg-mll}), constraining non-standard
physics parameters.

By attaching a~photon to the internal line of the loop one has an
effective neutrino--neutrino--photon vertex, which allows to calculate
the neutrino magnetic moment. In the case of Majorana neutrino, the CPT
theorem allows only for the transitional magnetic moment between two
neutrinos of different flavors. It is described by the effective
hamiltonian $H_{\rm eff} = \mu_{ii'} \bar\nu_{iL} (\sigma^{ab}/2)
\nu_{i' R}^c F_{ba}$, $F$ being the electromagnetic tensor.

After evaluating the loop amplitude one gets the following expression
for the magnetic moment:
\begin{equation}
  \mu_{ii'}^q = (1-\delta_{ii'}) \frac{3}{16\pi^2} m_{e^1} Q_d 
  \sum_{jk} \lambda'_{ijk} \lambda'_{i'kj} \left|
    \frac{2\sin(2\theta^k)}{m_{q^j}} g(x_2^{jk},x_1^{jk}) -
    (j \leftrightarrow k) \right |\mu_B,
\end{equation}
$m_{e^1}$ being the electron mass, $g(a,b)=(a\log(a)-a+1)/(1-a)^2 -
(b\log(b)-b+1)/(1-b)^2$, $Q_d=1/3$ is the down-quark electric charge,
and $\mu_B$ the Bohr magneton.  A~similar expression may be obtained for
the lepton--slepton contribution.

We have calculated the low energy MSSM spectrum by starting from certain
unification conditions at the GUT scale $\sim 10^{16}$ GeV and evolving
the running masses and coupling constants down to the $m_Z$ scale using
renormalization group equations (RGE). The so-obtained spectrum is
tested against various constraints on SUSY masses, proper electroweak
symmetry breaking, FCNC phenomenology and others. The procedure has been
described in detail elsewhere.\cite{mg-proc} We use the so-called
conservative approach, assuming that only one mechanism dominates at
a~time. It means that we perform calculations for some given $j,k$
without doing the summation. We pick up the highest obtained value,
which constitutes the ``most optimistic'' result:
\begin{eqnarray}
  \mu^q_{e\mu}       \le 4.0 \times 10^{-17} \mu_B, && \qquad \qquad
  \mu^\ell_{e\mu}    \le 1.6 \times 10^{-16} \mu_B,    \nonumber \\
  \mu^q_{e\tau}      \le 3.4 \times 10^{-17} \mu_B, && \qquad \qquad
  \mu^\ell_{e\tau}   \le 1.4 \times 10^{-16} \mu_B,              \\
  \mu^q_{\mu\tau}    \le 3.6 \times 10^{-17} \mu_B, && \qquad \qquad
  \mu^\ell_{\mu\tau} \le 1.4 \times 10^{-16} \mu_B.    \nonumber
\label{eq:mg-res}
\end{eqnarray}
These values were obtained for the following GUT conditions: $A_0=100$
GeV, $m_0 = m_{1/2} = 150$ GeV, $\tan(\beta) = 19$, $\mu>0$, where $A_0$
denotes the common value of soft breaking couplings, $m_0$ and $m_{1/2}$
the common scalar and fermion masses, and $\tan(\beta)$ is the ratio of
Higgs vacuum expectation values. By taking, for example, $A_0 \sim$ 1000
GeV, the results become smaller by at least one order of magnitude. We
would like to mention at this point that our bounds are around one order
of magnitude stronger than those published earlier.\cite{mg-loops} It is
due to the more exact procedure (GUT and RGE) and partially due to
inclusion of sparticle mixing. It is also worth mentioning that the
present experimental limits are $\mu_{ii'} \le 10^{-10}\mu_B$.

In further discussion one should analyze the influence of quark mixing
on $\mu_\nu$, which may give important corrections. This discussion is,
however, beyond the scope of the present work and will be postponed to
a~separate regular paper.

{\bf Acknowledgements }
This work was supported by the VEGA Grant agency of the Slovak Republic
under contract No.~1/0249/03, by the EU ILIAS project under contract
RII3-CT-2004-506222, and the Polish State Committee for Scientific
Research under grants no. 2P03B~071~25 and 1P03B~098~27.


\end{document}